\documentclass[
 superscriptaddress,
 reprint
]{revtex4-1}

\usepackage[utf8]{inputenc}
\usepackage{amsmath}    
\usepackage{graphicx}   
\usepackage{verbatim}   
\usepackage{color}      
\usepackage{subfigure}  
\usepackage{hyperref}   
\usepackage{wrapfig}
\DeclareUnicodeCharacter{00A0}{~}

\begin{document}

\title{Modulating the phase transition temperature of giant magnetocaloric thin films by ion irradiation}
\author{S.~Cervera}
\email[]{sophie.cervera@insp.jussieu.fr}
\affiliation{Institut des NanoSciences de Paris, Sorbonne Universités, UPMC Univ Paris 06, CNRS-UMR 7588,  75005, Paris, France}
\author{M.~Trassinelli}
\email[]{martino.trassinelli@insp.jussieu.fr}
\affiliation{Institut des NanoSciences de Paris, Sorbonne Universités, UPMC Univ Paris 06, CNRS-UMR 7588,  75005, Paris, France}
\author{M.~Marangolo}
\affiliation{Institut des NanoSciences de Paris, Sorbonne Universités, UPMC Univ Paris 06, CNRS-UMR 7588,  75005, Paris, France}
\author{C.~Carrétéro}
\affiliation{Unité Mixte de Physique, CNRS, Thales, Univ. Paris-Sud, Université Paris-Saclay, 91767, Palaiseau, France}
\author{V.~Garcia}
\affiliation{Unité Mixte de Physique, CNRS, Thales, Univ. Paris-Sud, Université Paris-Saclay, 91767, Palaiseau, France}
\author{S.~Hidki}
\affiliation{Institut des NanoSciences de Paris, Sorbonne Universités, UPMC Univ Paris 06, CNRS-UMR 7588,  75005, Paris, France}
\author{E.~Jacquet}
\affiliation{Unité Mixte de Physique, CNRS, Thales, Univ. Paris-Sud, Université Paris-Saclay, 91767, Palaiseau, France}
\author{E.~Lamour}
\affiliation{Institut des NanoSciences de Paris, Sorbonne Universités, UPMC Univ Paris 06, CNRS-UMR 7588,  75005, Paris, France}
\author{A.~Lévy}
\affiliation{Institut des NanoSciences de Paris, Sorbonne Universités, UPMC Univ Paris 06, CNRS-UMR 7588,  75005, Paris, France}
\author{S.~Macé}
\affiliation{Institut des NanoSciences de Paris, Sorbonne Universités, UPMC Univ Paris 06, CNRS-UMR 7588,  75005, Paris, France}
\author{C.~Prigent}
\affiliation{Institut des NanoSciences de Paris, Sorbonne Universités, UPMC Univ Paris 06, CNRS-UMR 7588,  75005, Paris, France}
\author{J.~P.~Rozet}
\affiliation{Institut des NanoSciences de Paris, Sorbonne Universités, UPMC Univ Paris 06, CNRS-UMR 7588,  75005, Paris, France}
\author{S.~Steydli}
\affiliation{Institut des NanoSciences de Paris, Sorbonne Universités, UPMC Univ Paris 06, CNRS-UMR 7588,  75005, Paris, France}
\author{D.~Vernhet}
\affiliation{Institut des NanoSciences de Paris, Sorbonne Universités, UPMC Univ Paris 06, CNRS-UMR 7588,  75005, Paris, France}



\begin{abstract}
Magnetic refrigeration based on the magnetocaloric effect at room temperature is one of the most attractive alternative to the current gas compression/expansion method routinely employed.
Nevertheless, in giant magnetocaloric materials, optimal refrigeration is restricted to the narrow temperature window of the phase transition ($T_c$).
In this work, we present the possibility of varying this transition temperature into a same giant magnetocaloric material by ion irradiation. 
We demonstrate that the transition temperature of iron rhodium thin films can be tuned by the bombardment of ions of Ne$^ {5+}$ with varying fluences up to 10$^{14}$ ions cm$^{-2}$, leading to optimal refrigeration over a large 270--380~K temperature window.
The $T_c$ modification is found to be due to the ion-induced disorder and to the density of new point-like defects. 
The variation of the phase transition temperature with the number of incident ions opens new perspectives in the conception of devices using giant magnetocaloric materials.
\end{abstract}

\maketitle

\section{Introduction}

The current climatic situation motivates investments in research and development of more environmentally friendly technologies.
In this context, the refrigerant systems based on the compression and expansion of an ideal gas, can be replaced by more efficient devices using materials which heat up or cool down when a magnetic field is applied or removed \cite{Franco2012, gschneidner_magnetocaloric_2000}.
This effect, named magnetocaloric effect, occurs in all magnetic materials and is more pronounced in the proximity of phase transitions due to the entropy variation involved.
Magnetocaloric materials with a phase transition near to room temperature have seen their interest growing up thanks to their promising properties in magnetic refrigeration \cite{brown_magnetic_1976, bruck_developments_2005, Franco2012, moya_caloric_2014, lyubina_magnetocaloric_2017} as well as in electricity generation \cite{kirol_numerical_1984-1, nielsen_review_2011, smith_materials_2012} for daily life applications (refrigerators, air and water conditioning, etc.).
A special attention has been recently devoted to magnetic materials that exhibit first-order phase transitions.
In such systems, the entropy variation and the associated capacity of the material to cool down are higher than for second order transitions.
As counterpart, this type of transition is generally characterized by an abrupt phase transformation that relegates the giant magnetocaloric effect in a small temperature window.
Paradoxically, materials exhibiting a large magnetocaloric effect, and then a high potential refrigeration capacity, are less suitable for realistic refrigerator devices that should work on several tens of degrees.

To overcome this issue, existing partially also in standard magnetocaloric materials, considerable efforts have been done in the past years. 
In particular, magnetocaloric materials with different transition temperatures ($T_c$) were combined in the form of heterostructures \cite{rowe_experimental_2006, dung_mixed_2011,liu_giant_2012, kuhn_magnetic_2011,christiaanse_generator_2014} to enlarge the working temperature interval of magnetic refrigerators.
In parallel, different methods to tune the transition temperature of giant magnetocaloric materials were investigated, among them, the variation of chemical composition and doping \cite{pecharsky_giant_1997,  pecharsky_tunable_1997,gschneidnerjr_recent_2005, de_campos_ambient_2006, rocco_ambient_2007, cui_magnetocaloric_2009, yue_structural_2015} but also the application of external strains \cite{liu_giant_2012, duquesne_ultrasonic_2012, moya_giant_2013}. 
The fabrication of suitable sets of giant magnetocaloric materials heterostructures requires however the \textit{ad hoc} synthesis of each individual component and a good thermal contact between them.

Here, we present an alternative method to tune the transition temperature based on the modification of giant magnetocaloric thin film properties using low velocity heavy ions.
More specifically, we demonstrate that changing the magnetic properties of epitaxial thin films of iron-rhodium (FeRh) is possible in a controlled manner by impact of ions.
FeRh, which is one of the most promising giant magnetocaloric materials, presents a magnetic phase transition of first order around $T_c$ = $375~K$ between an antiferromagnetic phase at low temperature and a ferromagnetic phase at high temperature.
During the transition, the pure ordered structure with B2 (Cs-Cl) symmetry does not change but the lattice volume increases of about 1\%.
At present, neither this high temperature of phase transition, nor the restricted window of the entropy variation ($\sim 20~K$) makes FeRh the best candidate for the magnetic refrigeration.
A variation of $T_c$ has been obtained in the past in non-equiatomic \cite{swartzendruber_ferh_1984} or doped FeRh \cite{kouvel_unusual_1966}.
In thin films of FeRh, only a $T_c$ increase is obtained by doping \cite{thiele_ferh/fept_2003,lu_first-order_2009}.
For all these cases the $T_c$ is modified to a given value on the entire sample and no gradient is achievable.
In the following, we report on the effects of heavy ion impact on the film properties, emphasizing the controlled modification of the transition temperature over a significant range. 
It is indeed the most relevant outcome for new conception of refrigeration devices.

\section{Experimental details}

\subsection{Growth conditions}
FeRh films with a $36$~nm thickness are grown by radio-frequency sputtering from a stoichiometric Fe$_{50}$Rh$_{50}$ target onto (001)-oriented MgO substrates \cite{cherifi_electric-field_2014}.
The growth is done at $630~^\circ$C with a power of $45$~W and under an argon pressure of $1\times10^{-3}$~mbar. 
After growth, the films are annealed in situ at $730~^\circ$C for $90$~min under vacuum (P~=~$1\times10^{-8}$~mbar).
The epitaxial films are then capped by $3$~nm of aluminum grown by dc sputtering at room temperature. 
The film thickness is determined by X-ray reflectivity.

\subsection{Irradiation conditions}

Ion irradiations are performed at an electron-cyclotron ion source, SIMPA facility (French acronym for multi-charged ion source of Paris, France) \cite{gumberidze_electronic_2010}.
The FeRh/MgO thin film obtained by sputtering was cut into six pieces.
Five of them are irradiated by Ne$^{5+}$ ions with a kinetic energy of $25$~keV and with an incident angle of $60^\circ$ (between the normal of the surface and the ion beam), while the last one was kept pristine and used as reference.
With these irradiation conditions, the ions have an average penetration equal to half the thickness of the FeRh layer and are deposited in all its volume.
At this energy, the nuclear elastic collisions between ions and FeRh atoms are the predominant processes \cite{Ziegler1985}.
They mainly induce point-like defect vacancies or interstitial atoms.
The number of those induced-defects can be controlled through the quantity of incident ions per cm$^{2}$, i.e. the fluence.
To investigate the possibility to tune the properties of FeRh thin films by varying the number of induced defects, the five samples are irradiated with different fluences from $2.8\times10^{12}$ to $1.1\times10^{14}$~ions~cm$^{-2}$ with an uncertainty $\leq8\%$. 
The number of ions per cm$^{2}$ impinging the targets is controlled by monitoring the ion beam intensity and the irradiation time.
The fluence is determined ``in-situ'' by using a position sensitive Faraday cup array \cite{panitzsch_direct_2009}, which provides the beam intensity profile, coupled to a CDD camera that takes images of the sample before and during irradiation.

\subsection{Magnetic and structural properties}
Magnetic properties of the different samples are determined with a superconducting interference device (SQUID, Quantum Design MPMS-XL 7 T) and a vibrating sample magnetometer (VSM, Quantum Design PPMS 9 T).
Before each measurement, the magnetic memory is deleted by applying a decreasing magnetic field oscillating around the zero average value.
Magnetization curves as a function of magnetic field are acquired for two temperatures $100$~K and $350$~K (the highest temperature reachable by this magnetometer) between $0$ and $~7$~T with steps of $0.01$~T.
The magnetization versus temperature curves are obtained with an applied magnetic field of $1$~T and with a sweep rate of $\pm~2$~K/min for all samples.

The irradiation effects on the FeRh layer structure are deduced by X-ray diffraction (XRD) measurements at room temperature (T~=~$293~\pm~1~K$) before and after irradiation (Rigaku Smartlab model).
The absence of sputtering induced by incident ions is checked with X-ray reflectivity technique.

\section{Results and discussion}
\begin{figure}
\centering
		\includegraphics[width=\columnwidth]{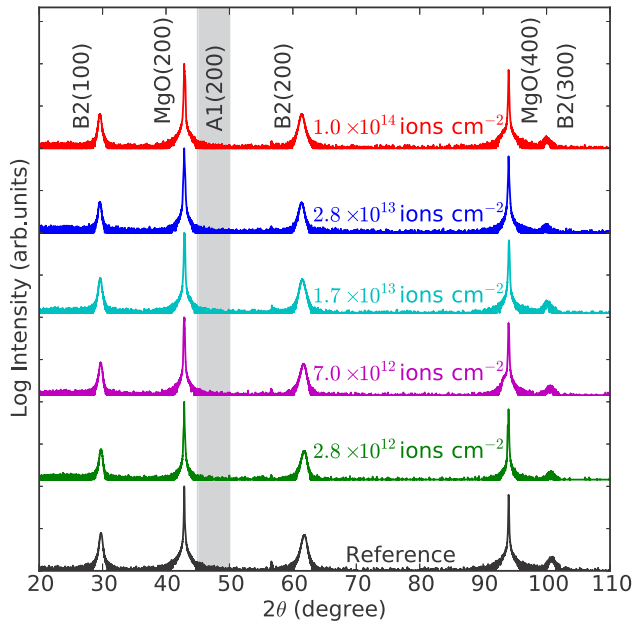}
		\caption{X-ray diffraction patterns ($\theta$-2$\theta$) for the reference and all irradiated samples submitted to different ion fluences.
The data are normalized to the (200) peak intensity of the MgO substrate for all the samples.}
		\label{fig:XRD}
\end{figure}

\begin{figure}
\centering
		\includegraphics[width=\columnwidth]{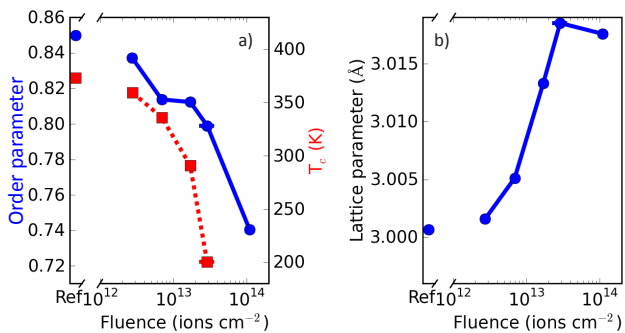}
		\caption{The order parameter (in lines), T$_{c}$ (in dashed lines) a) and the lattice parameter (in lines) b) of each samples as a function of the number of incident ions per cm$^{2}$.
The error bars are about the size of the symbols.}
		\label{fig:a_S_fct_flu}
\end{figure}

From the X-ray diffraction pattern of the different samples, displayed in Figure~\ref{fig:XRD}, several observations can be made.
The B$2$ (Cs-Cl) phase is clearly visible through the presence of three diffraction peaks whatever the ion fluence. Those peak intensities being barely affected, the irradiation does not amorphize the samples.
In previous works \cite{aikoh_study_2013, fujita_magnetic_2010, fujita_effects_2009}, the B$2$ phase was found to coexist with an A$1$-type phase (disordered face centered cubic structure) which is paramagnetic and expected to appear in the grey area in figure~\ref{fig:XRD}.
The absence of this phase in our XRD data attests the high quality of the films and the low damage induced by ion impact.
It means that those irradiations do not create a new crystalline phase, as an A$1$-type phase. 
Nevertheless, slight changes of the structural properties can be extracted from XRD. 
Indeed, the order parameter (figure~\ref{fig:a_S_fct_flu} a)) and the out-of-plane lattice parameter (figure~\ref{fig:a_S_fct_flu} b)) show a systematic variation with the ion fluence.
The order parameter, that is the fraction of Fe and Rh atoms in their cubic site \cite{vries_hall-effect_2013,Warren}, reflects the disorder induced by irradiation.
It can be calculated by: $s=\sqrt{I_{exp}(001)/I_{exp}(002)}/1.07$ where $I_{exp}(hkl)$ are experimental intensities and $1.07$ comes from theoretical intensities of the (hkl) Bragg reflection peak \cite{vries_hall-effect_2013}.
Figure~\ref{fig:a_S_fct_flu} a) shows a decrease of this parameter from $0.85$ to $0.73$.
Hence, the ion bombardment generates
disorder into the FeRh thin film leading to a slight modification of the lattice.
Indeed, the out-of-plane parameter gradually expands as a function of the irradiation fluence and reaches a relative expansion of $0.56$\% in the most irradiated sample (figure~\ref{fig:a_S_fct_flu} b)).
This expansion is attributed to the induced-disorder which enlarged the unit cell.

\begin{figure}
\centering
\includegraphics[width=\columnwidth]{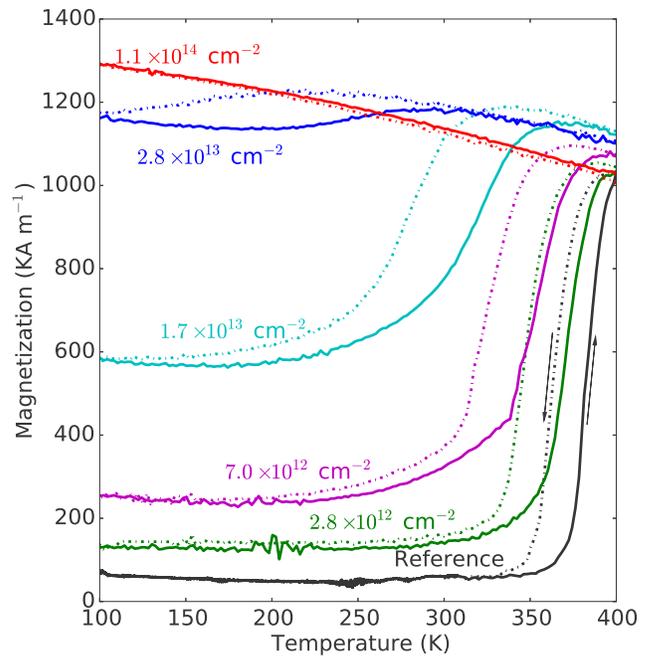}
\caption{The magnetization as a function of temperature for irradiated and reference samples. Data obtained, at 1 T, by
a temperature increase (solide lines) and decrease (dash-dotted lines).}
\label{fig:MT}
\end{figure}

The increase in disorder leads to the modification of the antiferromagnetic interaction between Fe atoms and changes the resulting sample magnetization.
Figure~\ref{fig:MT} represents magnetization as a function of temperature for different fluences.
First, it appears that irradiation induces a shift of the antiferromagnetic(AF)-to-ferromagnetic(F) transition temperature, as plotted on figure~\ref{fig:a_S_fct_flu} a) (dashed line).
Second, the magnetization at low temperature builds up as a function of the irradiation fluence (figure~\ref{fig:MT}) due to the creation of a stable and persistent F phase.
The robustness of the induced modifications has been tested repeating magnetization curves at different temperatures with a time interval of more than 2 years and after several temperature cycles up to 400~K without significant changes of the magnetic properties.
On the other hand, at high temperature, the magnetization does not change drastically, presenting only a slight increase as a function of the received ion fluence, followed by a decrease after a peak at $1.7\times10^{13}$~ions~cm$^{-2}$.

\begin{figure}
\centering
		\includegraphics[width=\columnwidth]{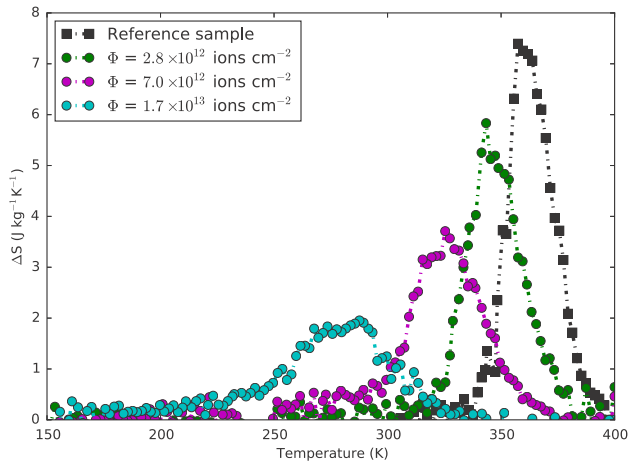}
		\caption{Entropy variation as a function of the temperature for the reference sample as well as three irradiated samples. The variation was determined from isofield magnetization curves for magnetic fields ranging from 0 to 2~T.}
		\label{fig:delta_S}
\end{figure}

It is worth noting that those irradiation-effects on the magnetic properties of FeRh thin films impact directly on their ability to refrigerate.
Actually, magnetocaloric materials are characterized by their cooling capacity, $q$.
This capacity is defined as the integration of the entropy variation $\Delta S$ within a fixed temperature interval.
Here, $\Delta S$ is the entropy variation generated by a 2~T magnetic field at a fixed temperature.
It is calculated from the sum of the derived curves of the magnetization versus decreasing temperature at various magnetic fields, applying the same procedure as in reference \cite{Mosca2008}, from $0.2$~T to $2$~T with steps of $0.2$~T.
The evolution of $\Delta S$ with temperature is shown in figure~\ref{fig:delta_S} for the sample irradiated with a fluence of $1.7\times10^{13}$~ions~cm$^{-2}$ and the reference one.
As expected, $\Delta S$ is maximal at the temperature of the phase transition, $290$~K for the irradiated sample and $375$~K for the reference one. 
The integration of the $\Delta S$ peak over the full width at half maximum gives the cooling capacity for the reference sample \cite{gschneidner_magnetocaloric_2000, smith_materials_2012}, $q$ = $144$~J~kg$^{-1}$, while for the irradiated one, $q$ = $84$~J~kg$^{-1}$.
The cooling capacity is then reduced by only a factor $2$ for a shift of temperature of $85$~K.
This reduction is mainly due to the enhancement of the magnetization at low temperature with the ion fluence.
\\

To enlighten the interpretation of our results, we recall what has been found in previous studies. 
The increase of magnetization at low temperature, as shown above, was already observed at 20 K 
in bulk FeRh and thin film samples (with thicknesses ranging from 0.2~mm to 80~nm) irradiated with energetic projectiles (ions and electrons) \cite{Fukuzumi2005,Kosugi2011,aikoh_study_2013, fujita_magnetic_2010, fujita_effects_2009, Zushi2007} and in FeRh thin films (with a thickness of 30 nm) irradiated with 30 keV Ga ions \cite{aikoh_quantitative_2011}.
Our results demonstrate that the disorder, induced by ion irradiation, is at the origin of the reduction of the AF exchange interaction. 
This reduction is more pronounced when the ion fluence increases, i.e. with the number of defects produced during ion collisional cascades. The possible role of ion implantation in the increase of magnetization can be excluded. Indeed, previous comparative experiments  \cite{aikoh_quantitative_2011} dedicated to the irradiation of FeRh thin films (30--80~nm) with fast (10~MeV iodine) and 
 slow ions (30~keV gallium), i.e. without and with ion implantation of the ions in the films respectively, show a similar behavior of the magnetization evolution with the fluence.
The AF phase is very sensitive to the local modification of the symmetry and periodicity of the lattice.
This is due to the variation of the local magnetic exchange terms leading to the stabilization of the F phase even at low temperature as experimentally observed \cite{schinkel_magnetic_1974, ohtani_features_1994} and theoretically predicted \cite{kaneta_theoretical_2011}.
Moreover, the observed shift in $T_c$ can also be related to the induced disorder.
Recently, theoretical study suggests that this shift is due to off-stoichiometry, namely Fe antisites that are connected to the defect density.
Those ab-initio calculations prove that slight off-stoichiometry induces a dramatic  drop in $T_c$ \cite{staunton_fluctuating_2014}, reflecting a compositionnal high sensitivity.
The relation between the defect density and the $T_c$ shift is also observed in experiments on FeRh thin films annealed at different temperatures \cite{cao_magnetization_2008}.
From our measurements we clearly show the strong correlation (figure~\ref{fig:a_S_fct_flu}) between the defect density (proportional to the fluence), the order parameter and the shift of transition temperature. 
This correlation unambiguously supports this theoretical interpretation.

\section{Conclusions and Perspectives} 

FeRh thin films of $36$~nm thickness have been irradiated with Ne$^{5+}$ ions at $25$~keV using different fluences. 
Ion impact modifies the properties of the magnetocaloric samples.
When varying the fluence from 0 to $10^{14}$~ions~cm$^{-2}$, we observe that the magnetization increases at low temperature.
This behavior is due to the progressive extinction of the antiferromagnetic interaction within the material showing that irradiation-induced defects promote the persistent ferromagnetic phase which does not participate to the phase transition.
Moreover, these defects amplify the disorder giving rise to a decrease of the transition temperature ($T_c$) due to a stabilization of the F phase at lower temperature.
The study, presented here on nanometric magnetocaloric materials allows us to highlight the mechanisms at the origin of the $T_c$ drop.
In particular, we demonstrate, for the first time, the existence of a clear correlation between the shift of $T_c$, the defect density and the induced disorder in the material.
Moreover, the relationship between $T_c$ and the ion fluence provides a new method to decrease and tune $T_c$, keeping an efficient refrigerant power in a temperature window about 120~K around room temperature.
The production of a single giant magnetocaloric thin film presenting different $T_c$ at different locations of the same material becomes possible thanks to this new technique \cite{Trassinelli2017a}.
Such sample could be readily produced with a spatial gradient of ion intensity or by modulating the irradiation time at different locations in the film.
It allows to get rid of the synthesis of different magnetocaloric materials and their subsequent critical assembly.
In particular, thin films with a spatial modulation of $T_c$ could be useful in cooling millimeter-electronic compounds for example or even in thermal energy harvesting devices \cite{ujihara_thermal_2007,Gueltig2017}.
Finally, in principle, this technique could also be generalized to bulk magnetocaloric materials, presenting a first order phase transition, for the development of efficient refrigerant or themomagnetic generator ``macro'' devices. 
In that respect further investigations are running using more energetic heavy ions to modify the transition temperature of micro- and millimetric samples, the typical size of powder grains and plates of active magnetocaloric materials in magnetic refrigerators.

\section{Acknowledgements}

Authors acknowledge M. LoBue and A. Bartok for their fruitful discussions, and Lee Phillips for his contribution.
This work was supported by French state funds managed by the ANR within
the Investissements d’Avenir programme under reference ANR-11-IDEX-0004-02, and
more specifically within the framework of the Cluster of Excellence MATISSE led by
Sorbonne Universit\'es.

\bibliography{bib_article_FeRh}

\end{document}